# A direct way to observe absolute molecular handedness

Zeev Vager

Department of Particle Physics and Astrophysics, Weizmann Institute, 76100 Rehovot, Israel

## Abstract

Unique labeling of chiral stereo-centers must include their handedness. The conventional method that has been developed to do this was originated by three chemists: R.S. Cahn, C. Ingold, and V. Prelog (CIP) and is formally known as R,S nomenclature. It requires knowledge of the spatial absolute configuration of that center. Traditionally, experimental methods of extracting handedness go through the absolute configuration and only then through the application of the CIP convention. Here we show that a direct experimental method of determination of the natural handedness by the polarization of tunneling electrons is almost always compatible with the CIP convention. By the sole use of symmetry arguments we show that the chiral molecule symmetry withdraws the need of fine structure splitting. As a consequence, the polarization of electrons tunneling through the molecular electric dipole direction uniquely determines their handedness. The symmetry breaking argument that induces lack of fine-structure splitting eliminates the need for a precise value of the spin-orbit interaction.

## Introduction

In stereochemistry, the convention of labeling chiral centers as "left" or "right" is given by Cahn–Ingold–Prelog priority (CIP) rules (1). These rules refer to the nuclear stereo-configuration of the center. Though it is a convention, each of the rules was chosen to convey profound chemical experience. The experimental determination of absolute chirality is always referred to this convention. The historical saga of the experimental efforts in such absolute determinations is well known (2). The experimental efforts are concentrated on finding the nuclear configuration of the examined molecules, which by itself is of foremost scientific importance. As such, the theory of quantum chemistry has been playing a vital role.

Accordingly if one could discover a relatively simple experimental procedure for finding the handedness of a given molecule, this would be a very worthwhile feat. In this note we show that this can be easily done by measuring the sign of the helicity of tunneling electrons. In addition, we will comment on the relation of the newly proposed method to the CIP convention. As remarked in (2) this is not applicable for molecules the gas phase.

The CIP algorithm for finding handedness can be described as follows. Given a carbon stereo-center, there are four bond directions which point to four imaginary spots on a sphere surrounding the center. The four bonding electrons of the carbon are shared with four groups on the imaginary sphere. Their density at the carbon center is zero. Essentially and except for minor details, the number of electrons on each group determines their directional priority. The highest priority is given to the group of highest number of electrons and so forth until the lowest. Crudely speaking, the electron distribution in the space around the carbon ion comes in four chemical groups. The number of electrons in each group defines the group priority. Usually, there is a dipole moment to such a distribution and its positive pole is in the direction of lowest priority group. Looking along the dipole direction, the CIP convention is that if the curved path connecting the groups along the descending priority turns around the direction of the dipole in a clockwise direction then it is a right handed center and vice versa. Notice that going from the positive pole to the negative pole along the increased priority curve offers the same clockwise rule for right handed centers and vice versa. We conclude from the above description that the path has the following properties.

1. It selects handedness.
2. It relates to the dipole direction by starting at one pole and ending at the other.
3. It chooses to pass through the most electron-populated volumes.

## The new idea

Considering a generic molecule we know that the Kramers degeneracy, applicable for odd number of electrons and no external magnetic field, allows for the spin polarization to be in any direction. A particular choice exclusive for chiral molecules (3) is along the dipole direction $\vec{d}$. For fixed nuclei bound chiral molecular state (4) there is no axis of symmetry, either discrete or continuous. Assuming that there is no spin-orbit (SO) interaction, the orbital angular momentum expectation value $\langle l_d \rangle$ must vanish. Upon introducing SO interaction, we expect $\langle l_d \rangle \neq 0$. This is valid for all chiral states. The sign of $\langle l_d \rangle$ is uniquely dependent on the spin direction since it is generated from zero by the spin. Therefore there is no meaning to fine-structure states in fixed nuclei bound chiral molecular states.

The non-existence of symmetry axis requires $\langle j_d \rangle = 0$ where $j = l + s$ for any size of SO interaction (except for the non-generic exact zero limit). Therefore, with spin polarization axis along (or opposite) the dipole of chiral molecular states

$$\langle l_d \rangle = -s_d = \pm \frac{1}{2}$$

Again, superposition of such pure spin polarization degenerate states is legitimate. This corresponds to the unusual appreciable orbital current within the chiral molecule which generates half a unit of magnetic flux.

It is also a demonstration of quantum anomaly. The orbital angular momentum and the spin angular momentum are entangled due to SO interaction. But, the constraint $\langle j_d \rangle = 0$ has no meaning when the spin-orbit interaction takes the limiting zero value. A measure of a free electron spin polarization which in its past interacted with the chiral state is highly correlated with the sign of $\langle l_d \rangle$.

The requirement $\langle l_d \rangle \neq 0$ necessitates orbital current. Since orbital current density is proportional to the electron density which is mainly localized on confined participating atoms, this unusual orbital current density is only locally circular. This is illustrated in Fig. 1. The above symmetry argument circumvents the need for formal quantum derivation of spin assisted current density, leading to the same conclusion for chiral bound states (5). Nevertheless, the concepts in that reference add to the intuitive understanding of many ideas introduced here and their connection to the SO interaction. The 3D extension of the formal quantum derivation of spin assisted current density is given in the Supplementary material.

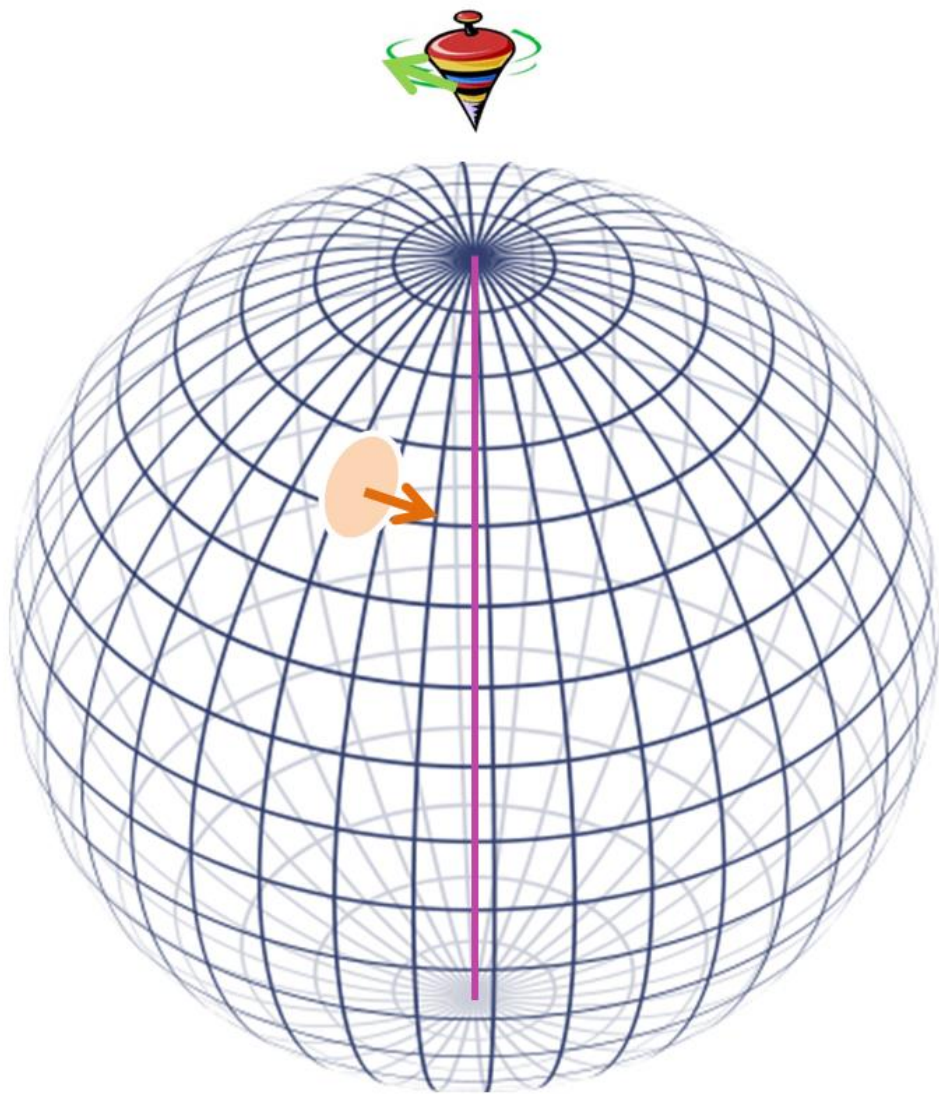

Fig. 1: The electron spin (above) along the dipole direction (purple) and orbital density current (brown arrow) attached to a local electron density on a molecular group within a chiral center.

## Definition of natural handedness

From the outset, quantum handedness is a symmetry breaking effect. It cannot relate to the pure isolated molecular Hamiltonian without additional environmental elements. Here, one of these elements is chosen to be the allowance of electron tunneling.

As shown above, the demand $\langle l_d \rangle = -s_d = \pm \frac{1}{2}$ demonstrates a spin dependent orbital current around the dipole for bound chiral states. Imagine converting the bound state into a narrow resonant state by tunneling of an electron which penetrates the positive pole and exits the negative pole. Now, the necessarily continuous orbital current is composed of circular+translational motion which corresponds to a helical handed motion. Whichever direction is the tunneling along the dipole is, $\langle l_d \rangle > 0$ defines a right handed orbital current. Correspondingly, $\langle l_d \rangle < 0$ defines a left handed orbital current.

If the physics of tunneling through chiral molecules where to proceed only through such selective single handed tunneling then the natural handedness nomenclature should have been in accordance with the allowed handedness. Right handed when $\langle l_d \rangle > 0$ and left handed when $\langle l_d \rangle < 0$. Such currents populate only one of the Kramers twin states and therefore the steady state involves the pure transmission of $s_d = \mp \frac{1}{2}$ in accordance with the path being right or left handed.

Indeed the most probable tunneling through molecular insulators is via very narrow molecular resonances. The molecular dipoles induce electric fields in their environments such that the tunneling electron penetrates the positive pole and exits at the negative pole. Therefore we adopt the above definition of handedness as the natural one.

## Connection to the CIP convention

We now come back to the listed properties of the CIP convention. Since a current through a molecule needs a continuous density along its path, the priority selected paths are preferred. The CIP handed path defines a sign to $\langle l_d \rangle$ and selectively ($\langle j_d \rangle = 0$) populates the spin of the molecule along the dipole direction. The steady state tunneling transmits pure helicity electrons. Thus, there is a simple way to measure the handedness of a chiral species without the knowledge of the absolute configuration. All what is required is the mere existence of an electrical dipole and a measure of a tunneling electron spin helicity.

## Discussion

Spin polarization of tunneling electrons through chiral molecules and related experiments have been carried regularly for more than a decade at the Weizmann-Institute Chemical-Physics laboratory of Ron Naaman's group (6). It is believed that the theoretical arguments of this article contain a firm basis to the robustness of the spin selectivity and its relation to handedness. Other attempts to explain the spin-selectivity (7) are mainly based on modeling the SO interaction for specific chiral systems and are believed here to miss the essence of the symmetry argument in this article. In particular, the SO-interaction independence of the chiral $\langle j \rangle = 0$ argument for chiral systems is vital. The need for a dipolar axis for the spin selectivity emerges from the necessity to define tunneling direction. Lack of a dipole allows tunneling along directions which do not follow order such as the priority list of CIP. This is easily remedied by external electric field polarization.

A closer comparison between the CIP priorities and the corresponding odd electron priority list among the participating molecular groups reveals possible differences. The choice of the natural handedness is connected to the odd electron distribution. It is the odd electron which is mainly responsible for charge displacement, creating the dipole as the pivot of the helical priority distribution.

Since the understanding of chiral stability is through interaction with the environment, the actual number of electrons belonging to the molecule is a fluctuating quantity. Mainly unpaired electrons contribute to the spin selectivity and charge displacement; therefore, the spin selected tunneling is not always the main tunneling process. Nevertheless, the sign of the tunneling-electron helicity always reveals the handedness of the species.

## Supplementary material: 3D spin assisted current

The Rashba Hamiltonian is

$$H = -\frac{\hbar^2}{2m}\nabla^2 + i(\vec{F} \times \vec{\nabla}) \cdot \vec{S} + V$$

where $\vec{F}$ is proportional to the electric field $-\vec{\nabla}V$.

The time dependent equation and its complex conjugate are

$$\Psi^\dagger \left( i\hbar \partial_t + \frac{\hbar^2}{2m}\nabla^2 - i(\vec{F} \times \vec{\nabla}) \cdot \vec{S} - V \right)\Psi = 0$$

$$\Psi^T \left( -i\hbar \partial_t + \frac{\hbar^2}{2m}\nabla^2 + i(\vec{F} \times \vec{\nabla}) \cdot \vec{S}^* - V \right)\Psi^* = 0$$

Upon subtraction

$$i\hbar \partial_t \rho + \frac{\hbar^2}{2m}\left( \Psi^\dagger \nabla^2 \Psi - \Psi^T \nabla^2 \Psi^* \right) - i\left( \Psi^\dagger (\vec{F} \times \vec{\nabla}) \cdot \vec{S}\Psi + \Psi^T (\vec{F} \times \vec{\nabla}) \cdot \vec{S}^* \Psi^* \right) = 0$$

$$\frac{\hbar^2}{2m}(\Psi^\dagger \nabla^2 \Psi - \Psi^T \nabla^2 \Psi^*) = \frac{\hbar^2}{2m}\vec{\nabla} \cdot (\Psi^\dagger \vec{\nabla}\Psi - \Psi^T \vec{\nabla}\Psi^*) = i\hbar \vec{\nabla} \cdot \frac{i\hbar}{2m}(\Psi^T \vec{\nabla}\Psi^* - \Psi^\dagger \vec{\nabla}\Psi)$$

$$\equiv i\hbar \vec{\nabla} \cdot \vec{I}_O$$

We define four real functions by

$$\Psi \equiv \begin{pmatrix} \Psi_+ \\ \Psi_- \end{pmatrix} = \begin{pmatrix} \sqrt{\rho_+}e^{i\Phi_+} \\ \sqrt{\rho_-}e^{i\Phi_-} \end{pmatrix}$$

The orbital current density $\vec{I}_O$ and the spin assisted current density $\vec{I}_S$ are given by

$$\vec{I}_O = \frac{\hbar}{m}(\rho_+\vec{\nabla}\Phi_+ + \rho_-\vec{\nabla}\Phi_-)$$

$$\vec{I}_S = \vec{S} \times \vec{F}$$

where

$$\vec{S} \equiv \Psi^\dagger \vec{S} \Psi$$

Therefore the total probability current density is $\vec{I} = \vec{I}_O + \vec{I}_S$.

If the spin is polarized along the z axis then

$$\vec{I}_O = \frac{\hbar}{m}\rho_+\vec{\nabla}\Phi_+ \qquad \vec{I}_S = \rho_+\vec{e}_z \times \vec{F}$$

The above is presented with the permission of the originator – Dan Vager.

For bound states the orbital density current is simply equal to $-\vec{I}_S$. Use of the Hellmann–Feynman (8) theorem provides us the force on the local odd electron density. In particular, the force is mainly towards the center of positive charge. When the locality is away from the z axis then $\vec{I}_O = -\vec{I}_S$ is non-zero in the direction shown in Fig. 1.

## References and notes

1. R.S. Cahn, C. Ingold, and V. Prelog, Specification of Molecular Chirality. *Angew. Chem. Int. Ed.* **5** (4): 385–415 (1966).
2. Absolute configurations for a chiral molecule are most often obtained by *X-ray crystallography*. Special technique must be used for the gas phase as in: Philipp Herwig *et al*, Imaging the Absolute Configuration of a Chiral Epoxide in the Gas Phase, *Science* **342**, 1084 (2013) and Kerstin Zawatzky *et al*, Coulomb Explosion Imaged Cryptochiral (R,R)-2,3-Dideuterooxirane, *Chem. Eur. J.* **20**, 5555 (2014).
3. The projection of angular momentum on the dipole direction is a pseudo-scalar. Its value is zero for parity conserving states.
4. Differently from the gas phase, the existence of environment is always assumed. Thus, symmetry broken phenomena, such as chiral structure and dipole direction are implicit. Accordingly, the symmetry of the molecular odd electron states is affected.
5. Dan Vager, Zeev Vager, Spin order without magnetism – a new phase of spontaneously broken symmetry in condensed matter, *Phys. Lett. A* **376,** 1895–1897 (2012).
6. Ref. (7) provides a good review of the relevant experimental publications.
7. Joel Gersten, Kristen Kaasbjerg, Abraham Nitzan, Induced spin filtering in electron transmission through chiral molecular layers adsorbed on metals with strong spin-orbit coupling, *J. Chem. Phys.* **139**, 114111 (2013).
8. Feynman, R. P. Forces in Molecules. *Phys. Rev.* **56** (4): 340 (1939).

## Acknowledgements

Helpful discussion and suggestions by Prof. Itamar Procaccia are highly appreciated. The 3D spin assisted current formalism in the supplementary material was graciously provided by Dan Vager, the coauthor in ref. (5).